\documentclass[conference]{IEEEtran}
\IEEEoverridecommandlockouts
\usepackage{cite}
\usepackage{amsmath,amssymb,amsfonts}
\usepackage{algorithmic}
\usepackage{graphicx}
\usepackage{textcomp}
\usepackage{xcolor}

\definecolor{mygreen}{rgb}{0,0.6,0}

\usepackage{booktabs}
\usepackage{tikz}

\usepackage{makecell}
\usepackage{multirow}
\usepackage{adjustbox}
\usepackage{subcaption}
\usepackage{enumitem}
\usepackage{comment}

\newcommand{\squeezeup}{\vspace{-0mm}}

\usepackage[acronym]{glossaries}

\usepackage[colorinlistoftodos]{todonotes}

\def\BibTeX{{\rm B\kern-.05em{\sc i\kern-.025em b}\kern-.08em
		T\kern-.1667em\lower.7ex\hbox{E}\kern-.125emX}}

\newcommand{\anonymize}[1]{\textit{(removed for review)}}

\newcommand{\hide}[1]{}

\begin{document}


\newacronym{zkp}{ZKP}{Zero-knowledge Proofs}
\newacronym{nizk}{NIZK}{non-interactive zero-knowledge proof}
\newacronym{gdpr}{GDPR}{General Data Protection Regulation}
\newacronym{ip}{IP}{Interactive Proofs }
\newacronym{p2p}{P2P}{Peer-to-Peer}
\newacronym{voc}{VOC}{Verifiable Off-chain Computation}
\newacronym{zksnark}{zk-SNARK}{zero-knowledge succinct non-interactive argument of knowledge}
\newacronym{cli}{CLI}{Command Line Interface}
\newacronym{iot}{IoT}{Internet of Things}
\newacronym{iac}{IaC}{Infrastructure as Code}
\newacronym{api}{API}{application program interface}
\newacronym{nist}{NIST}{National Institute of Standards and Technology}
\newacronym{mb}{Mb}{megabyte}
\newacronym{zkstark}{zk-STARK}{zero-knowledge Scalable Transparent Argument of Knowledge}
\newacronym{dsl}{DSL}{Domain-Specific Language}
\newacronym{soa}{SOA}{Service-Oriented Architecture}
\newacronym{mpc}{MPC}{Multi-party Computation}
\newacronym{abi}{ABI}{Application Binary Interface}
\newacronym{gb}{GB}{gigabytes}
\newacronym{plonk}{PLONK}{Permutations over Lagrange-bases for Oecumenical Noninteractive arguments of Knowledge)}
\newacronym{tps}{TPS}{transaction per seconds}
\newacronym{ivc}{IVC}{Incremental Verifiable Computation}
\newacronym{nfs}{NFS}{Network File System}
\newacronym{ecs}{$ecs$}{Executable Constraint System}
\newacronym{zkvm}{zk-VM}{Zero-knowledge Virtual Machine}

	\title{Servicifying zk-SNARKs Execution for \\Verifiable Off-chain Computations
		{\footnotesize }
		\thanks{}
        \squeezeup
	}
	
	\author{\IEEEauthorblockN{Alvaro Alonso Domenech, Jonathan Heiss, Stefan Tai}
		\IEEEauthorblockA{\textit{Information System Engineering } \\
			\textit{TU Berlin}\\
			Berlin, Germany \\
			{\{aa,jh,st\}}@ise.tu-berlin.de}
	}

	\maketitle	

	\begin{abstract}
         Zk-SNARKs help scale blockchains with Verifiable Off-chain Computations (VOC). 
         \acrshort{zksnark} DSL toolkits are key when designing arithmetic circuits but fall short of automating the subsequent proof-generation step in an automated manner.
         We emphasize the need for portability, interoperability, and manageability in \acrshort{voc}-based solutions and introduce a Proving Service that is designed to provide a scalable and reusable solution for generating \acrshort{zksnark} proofs leveraging clouds.
	\end{abstract}
	
	
	\section{Introduction}
	\label{sec:introduction}
	
The succinctness property and the short verification times of \acrshort{zksnark}s come at the cost of large computational complexity, static execution models, and memory overheads during the proof generation process~\cite{memory_intensity_proof_generation}.
These limitations represent a problem in VOC applications which require handling large and varying workloads, like rollups. 
This type of application would benefit from scalable, interoperable, and manageable system environments like clouds.
However, DSL toolkits like ZoKrates~\cite{paper_eberhardt_zokrates} or Circom~\cite{circom_paper} concentrate on circuit development and put little emphasis on how to integrate these circuits into production systems.
Bridging the gap between advanced cryptography and modern systems engineering also helps to create standardized benchmarks for proving systems and tools.

	\section{Proving Service}
	\label{sec:paas}
	We introduce a service-oriented approach for VOC that facilitates the use of the cryptographic procedures of \acrshort{zksnark}s within cloud system architectures. 
Our system allows for executing arithmetic circuits as encapsulated application logic in containers that are deployable to different machines realizing scalability, provide interoperability with other services, and are better manageable. 
For that, we give an overview of the service-oriented system architecture and describe the internals of the proving service.

\subsection{Service Oriented Verifiable Off-chain Computation}
\begin{figure}[t!]
    \centering
    \includegraphics[width=0.8\columnwidth]{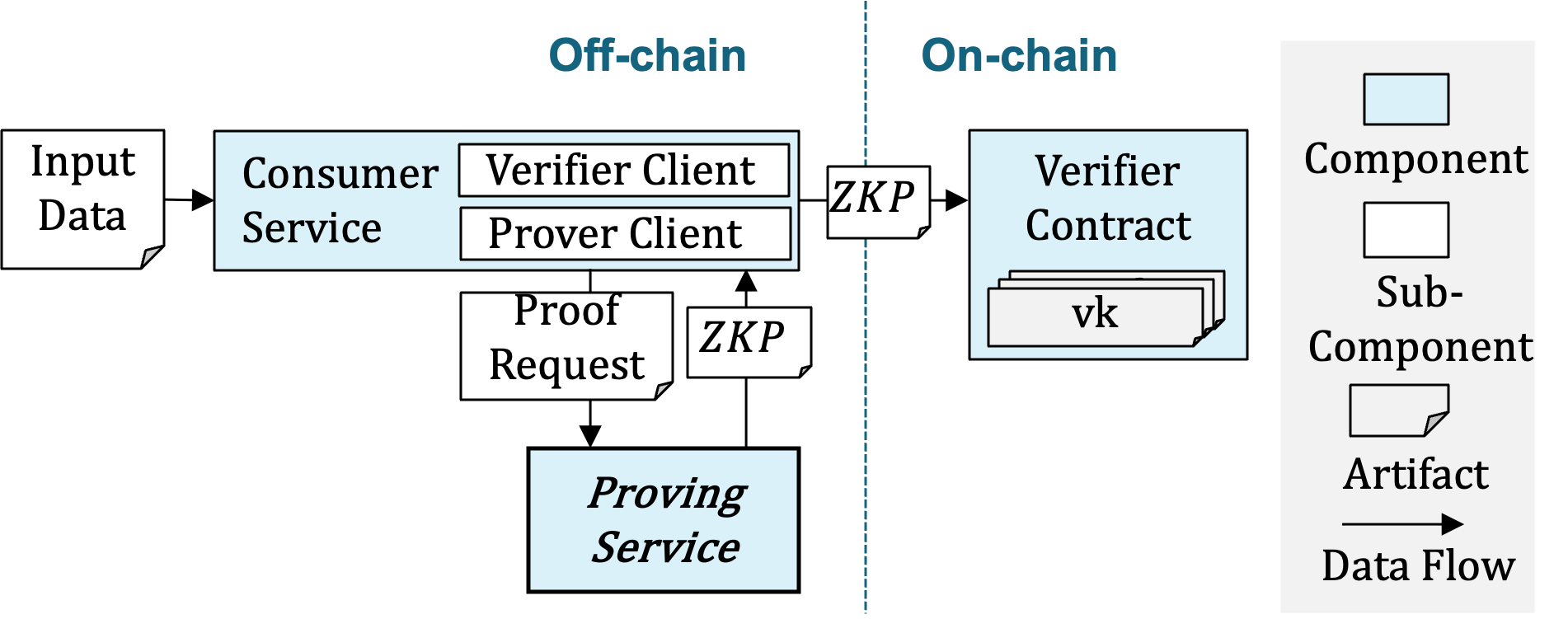}
    \caption{Proving Service Model}
    \label{fig:system_overview}
\end{figure}

The problems above can be fulfilled through a service-oriented approach as depicted in Figure~\ref{fig:system_overview}.
Starting from a higher-level system's perspective, we treat the proving service as a black box which, upon a request, 
returns the proof together with the computation's output.
Following the \acrshort{voc} model~\cite{paper_eberhardt_zokrates}, we distinguish between the blockchain infrastructure hosting the \textit{verifier contract} and a cloud-based off-chain infrastructure that runs outside the consensus protocol and hosts the \textit{consumer} and the \textit{proving service}. 

The \textit{consumer service} is responsible for managing the proving service's inputs and outputs and interacting with the relying verifier contract. 
It receives data from external sources and translates them into a \textit{proof request}. 
Upon a request through the \textit{prover client}, the proving service executes the \acrshort{voc} 
and returns the \textit{ZKP} attesting to the computational integrity of the VOC. 
The consumer service then submits the ZKP to the verifier contract through its \textit{verifier client}. 
On submission, the verifier contract verifies the proof computed by the proving service. 

\begin{figure}[bth]
    \centering
    \includegraphics[width=0.8\columnwidth]{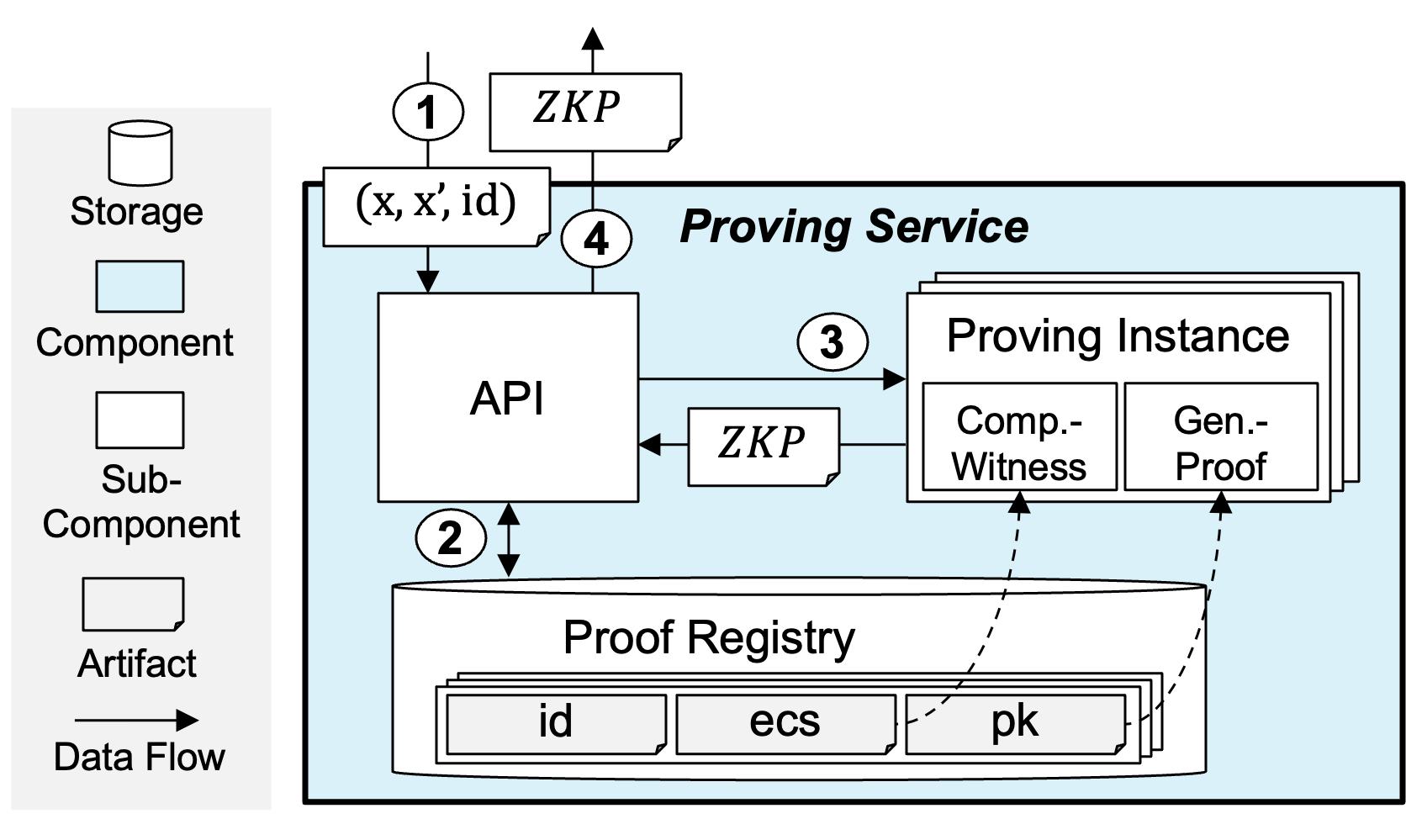}
    \caption{Proving Service Architecture}
    \label{fig:proving_service_architecture}
\end{figure}

\begin{figure*}[htb!]
    \centering
    \includegraphics[width=1.5\columnwidth]{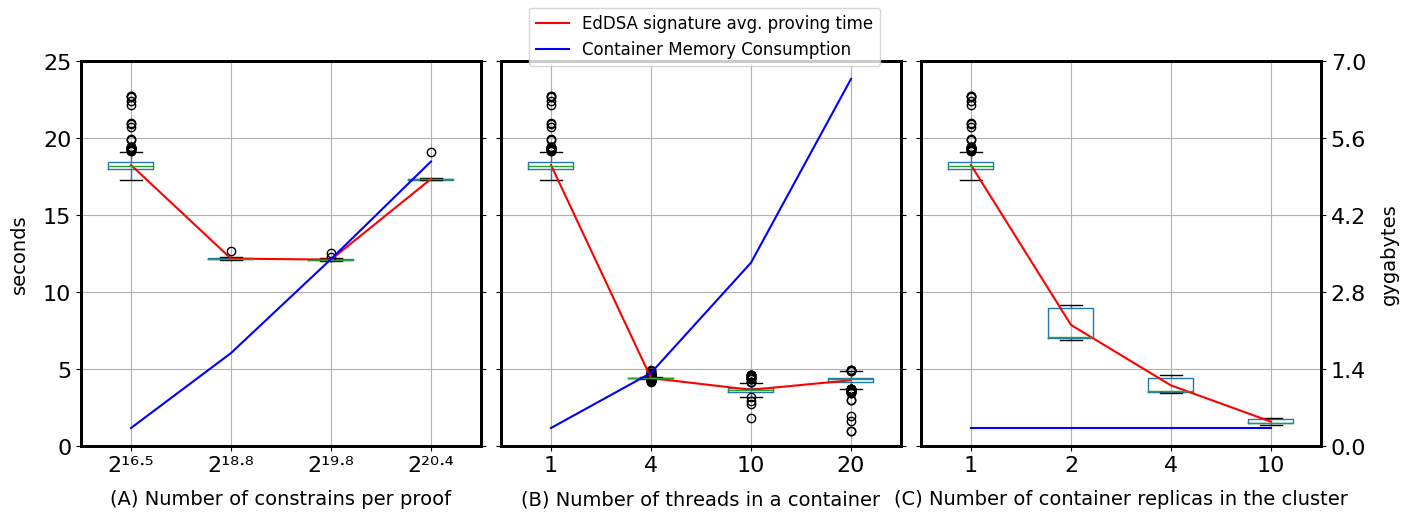}
    \caption{Proving Time and Memory Consumption}
    \label{fig:experiments_results}
\end{figure*}

\subsection{Service Architecture}
The \textit{Proving Service} depicted in Figure~\ref{fig:proving_service_architecture} is an application service that runs stand-alone on the prover's off-chain infrastructure. 
It serves proof requests by exposing the proving-related operations through an \textit{Application Program Interface} (API) as minimal functionality to the consumer service.

The procedure can be summarized in four simple steps:
First, the proving service receives the \textit{Proof Request} containing the public ($x$) and private ($x'$) proof arguments and the identifier of the addressed \acrfull{ecs} ($id$).
Second, using the $id$ the API fetches the corresponding $ecs$ and \textit{private key} ($pk$) from the \textit{Proof Registry} which is the persistent storage component containing these large, recurrently requested files needed for proving.
The proof registry manages different reusable pairs of $ecs$ and $pk$, each addressable through a unique $id$.

Third, the \textit{Proving Instance} is executed in two stages: the witness computations and the proof generation.
Fourth, the ZKP is returned to the consumer service through the API upon successful execution.  

    \section{ZoKrates-API}
	\label{sec:techn_eval}
	
The previous service-oriented architecture serves as a technology-agnostic blueprint for building proving services for DSL circuits. 
For evaluation, we technically instantiate the proving service for ZoKrates~\cite{paper_eberhardt_zokrates} and present the ZoKrates-API\footnote{https://github.com/ZK-Plus/zokrates-api}
as a ready-to-use open-source software. 
We servicify ZoKrates by wrapping an API around the ZoKrates interpreter, the central component of the ZoKrates software that previously has only been addressable through a Command Line Interface and a Javascript library. 
The ZoKrates-API exposes the methods of the ZoKrates interpreter through HTTP endpoints.
We containerized the ZoKrates-API using Docker, making the services easily deployable among a wide range of machines and allowing us to further leverage cloud-native tools like Kubernetes for horizontal scalability, manageability, and observability. 
Furthermore, the ZoKrates-API supports multi-threading so a single instance can compute multiple proofs in parallel.

\section{Evaluation}

To test the presented implementation, we deployed the containerized ZoKrates-API on a Kubernetes cluster.
As a workload for our experiments, we generated a large number of EdDSA signatures\cite{heiss_trustworthyPreprocessing_2021} which amounts for more than $2^{29}$ of circuit constraints similar to \cite{chiesa_eos}.

We conducted three experiments (see Figure \ref{fig:experiments_results}) to measure the average proving time per signature in [sec] and the memory consumption in [gb] using various cluster configurations. 
Figure~\ref{fig:experiments_results}A) shows a 33\% improvement in proving time for a single machine when choosing an appropriate machine size. 
An argument in favour of vertical escalability.
For a single machine as well, Figure~\ref{fig:experiments_results}B) shows that the proving time can be brought down drastically when few parallel threads are enabled, though the gains plateaued rapidly due to the increasing resources needed. 
Running the same experiments in parallel VMs instead of threads, Figure~\ref{fig:experiments_results}C) demonstrate a better approach to scaling proving as the computational burden is distributed over several machines, proportional increasing the processing time.

\section{conclusion}

As the experiments demonstrate, significant performance improvements can be gained from horizontal (more nodes) and vertical (larger nodes) scalability of an arbitrary \acrshort{zksnark} proof. 
We facilitate this process by leveraging modern cloud-native architectures such as Docker and Kubernetes.

	\bibliographystyle{IEEEtran}
	\bibliography{references}

\end{document}